\documentclass[%
reprint,
twocolumn,
notitlepage,
superscriptaddress,
amsmath,amssymb,
 pra,
]{revtex4-2}
\usepackage{amsmath}
\usepackage{graphicx}
\usepackage{subfigure}
\usepackage{dcolumn}
\usepackage{bm}
\graphicspath{{figure/}}
\usepackage{array}
\usepackage{braket}
\usepackage{diagbox}
\usepackage{appendix}
\usepackage{multirow}
\usepackage{bm,color,bbm}
\usepackage{booktabs}

\newcolumntype{.}{D{.}{.}{-1}}
\usepackage{amsopn}

\usepackage{epstopdf}

\usepackage{makecell}

\usepackage[percent]{overpic}  

\usepackage{threeparttable}
\usepackage{siunitx}
\usepackage{placeins}
\usepackage{longtable}
\usepackage[colorlinks,linkcolor=blue, anchorcolor=blue, urlcolor=blue, citecolor=blue]{hyperref}
\usepackage{cleveref}

\begin{document}
\title{High-rate  quantum digital signatures over 250 km of optical fiber}

\author{Jiemin Lin}
\author{Yongqiang Du}
\author{Mingxuan Zhang}
\author{Ruiheng Jing}
\author{Xin Liu}
\affiliation {Guangxi Key Laboratory for Relativistic Astrophysics, School of Physical Science and Technology, Guangxi University, Nanning 530004, China}

\author{Xiaodong Liang}
\affiliation {Guangxi Key Laboratory of Optical Network and Information Security, The 34th Research Institute of China Electronics Technology Group Corporation, Guilin 541004, Guangxi, China.}

\author{Hongbo Xie}
\author{Yanwei Li}
\affiliation {Ji Hua Laboratory, Foshan 528200, China.}

\author{Hua-Lei Yin}
\affiliation {Department of Physics and Beijing Key Laboratory of Opto-electronic Functional Materials and Micro-nano Devices, Key Laboratory of Quantum State Construction and Manipulation (Ministry of Education), Renmin University of China, Beijing 100872, China.}

\author{Kejin Wei}
\affiliation {Guangxi Key Laboratory for Relativistic Astrophysics, School of Physical Science and Technology, Guangxi University, Nanning 530004, China}

\begin{abstract}	

Quantum digital signatures (QDS) offer information-theoretic security for message integrity, authenticity, and non-repudiation, and constitute a fundamental cryptographic primitive for future quantum networks. Despite significant progress, the practical deployment of QDS has been severely constrained by limited signature rates and poor tolerance to channel loss, particularly in long-distance and metropolitan-scale networks. Here, we report a high-rate, loss-resilient QDS system that overcomes these two key bottlenecks simultaneously. Our implementation combines intrinsically phase-stable polarization modulation based on a Sagnac interferometer with gigahertz-rate quantum state encoding and low-timing-jitter superconducting nanowire single-photon detectors, enabling robust and continuous operation at high repetition frequencies. By integrating this hardware platform with a one-time universal hashing-based QDS protocol, we achieve a signature rate improvement of more than two orders of magnitude compared with existing QDS implementations under comparable channel-loss conditions. Notably, the system maintains a non-zero effective signature rate of approximately 1.25 times per second at a total channel loss of up to 49.05 dB, representing the highest loss tolerance reported for QDS to date. These results establish a practical and scalable technological pathway for deploying QDS in real-world quantum communication networks.
\end{abstract}

\maketitle

\section{Introduction}

With the maturation of quantum key distribution (QKD) as the first generation of quantum network technology, secure key establishment over metropolitan and intercity fiber networks has become increasingly reliable and deployable~\cite{kimble2008quantum,castelvecchi2018quantum,wehner2018quantum,singh2021quantum,li2023entanglement,delle2025operating}. As QKD transitions from laboratory demonstrations~\cite{wang2022twin,grunenfelder2023fast,wei2023resource,liu2023experimental,liu2024reference,liu2026high} toward real-world infrastructures~\cite{chen2021twin,zhu2024field,zahidy2024quantum,zhou2024independent,wu2025integration,pittaluga2025long,guan2025field}, the focus of quantum communication is shifting from secure key generation alone to higher-layer cryptographic functionalities that enable complex, accountable, and trustworthy interactions in quantum networks. In particular, emerging network-scale applications-such as digital contracts, financial transactions, and distributed ledgers-require security guarantees that go beyond confidentiality, including message authenticity, integrity, and non-repudiation.

To meet these emerging requirements beyond secure key distribution, QDS have been proposed as a fundamentally new cryptographic primitive for quantum networks. QDS provide information-theoretic security guarantees for message authenticity, integrity, and non-repudiation, enabling trustworthy communication among distributed parties. In contrast to classical digital signature schemes, whose security relies on computational hardness assumptions~\cite{rivest1978method,elgamal1985public,johnson2001elliptic}, QDS derives its security directly from the laws of quantum mechanics, and therefore remains secure even against adversaries with unbounded computational power, including future large-scale quantum computers~\cite{arute2019quantum,arrazola2021quantum,bluvstein2024logical,gao2025establishing}.

Since the seminal proposal by Gottesman and Chuang in 2001~\cite{gottesman2001quantum}, QDS have undergone a sustained evolution from an idealized theoretical construct to a practically implementable cryptographic primitive. Early QDS protocols relied on demanding assumptions, including long-lived quantum memory and fully trusted quantum channels, which rendered experimental realization infeasible. Subsequent advances in protocol design and security analysis progressively removed these constraints~\cite{clarke2012experimental,dunjko2014quantum,yin2016practical,amiri2016secure}, enabling QDS implementations under realistic system conditions~\cite{yin2017experimental102,roberts2017experimental,ding2020280}. 

A major milestone in this evolution is the introduction of the one-time universal hashing quantum digital signature (OTUH-QDS) protocol~\cite{yin2023experimental,li2023one}, which enables efficient, information-theoretically secure signing of multi-bit messages without sacrificing security. By decoupling signature efficiency from message length while retaining unconditional security, OTUH-QDS establishes a favorable balance between security, scalability, and implementation complexity. This protocol-level breakthrough provides a solid foundation for advancing QDS from proof-of-principle demonstrations toward high-performance, application-oriented quantum signature systems.

Building upon these protocol-level advances, experimental QDS platforms have progressed from proof-of-principle demonstrations toward increasingly sophisticated implementations, incorporating higher repetition rates~\cite{collins2016experimental,collins2017experimental,an2018practical,richter2021agile}, more complex network topologies~\cite{pelet2022unconditionally,lu2025fully}, and demonstrations in both field-deployed fiber links~\cite{yin2017experimental,chapman2024entanglement} and chip-integrated platforms~\cite{du2025chip}. Despite these developments, the practical deployment of QDS remains fundamentally constrained by two tightly coupled system-level bottlenecks. First, achieving high signature rates requires quantum state preparation and detection at gigahertz repetition frequencies, which imposes stringent demands on long-term phase stability, modulation fidelity, and single-photon detector performance~\cite{hadfield2009single}. Second, realistic fiber networks are intrinsically lossy, and increasing channel loss rapidly suppresses the effective photon detection probability, leading to a severe degradation of signature throughput. As a result, existing QDS systems typically face a trade-off between high-rate operation and tolerance to high channel loss. Overcoming this trade-off and enabling simultaneously high-rate and loss-resilient QDS operation remains a central challenge for scalable QDS.

In this work, we address this challenge by demonstrating a QDS system that combines high-rate operation with strong tolerance to channel loss. Our implementation integrates intrinsically stable polarization modulation based on a Sagnac interferometer~\cite{li2019high,ma2021simple,luo2022intrinsically} with gigahertz-rate quantum state encoding and low-timing-jitter superconducting nanowire single-photon detectors (SNSPDs), enabling robust and continuous operation at high repetition frequencies. Combined with the OTUH-
QDS protocol, the system supports efficient signing of messages of arbitrary length. Experimentally, we achieve a signature rate of 5186.80 times per second (tps) over a 75~km standard single-mode fiber link. Remarkably, the system maintains a non-zero effective signature rate of approximately 1.25~tps under a total channel loss of up to 49.05~dB, corresponding to a fiber length of 250~km. These results establish a practical and scalable technological pathway toward the deployment of QDS in real-world quantum communication networks.

\section{Protocol}

\begin{figure*}[t]
	\begin{center}
		\begin{tabular}{c}
			\includegraphics[height=8.5cm]{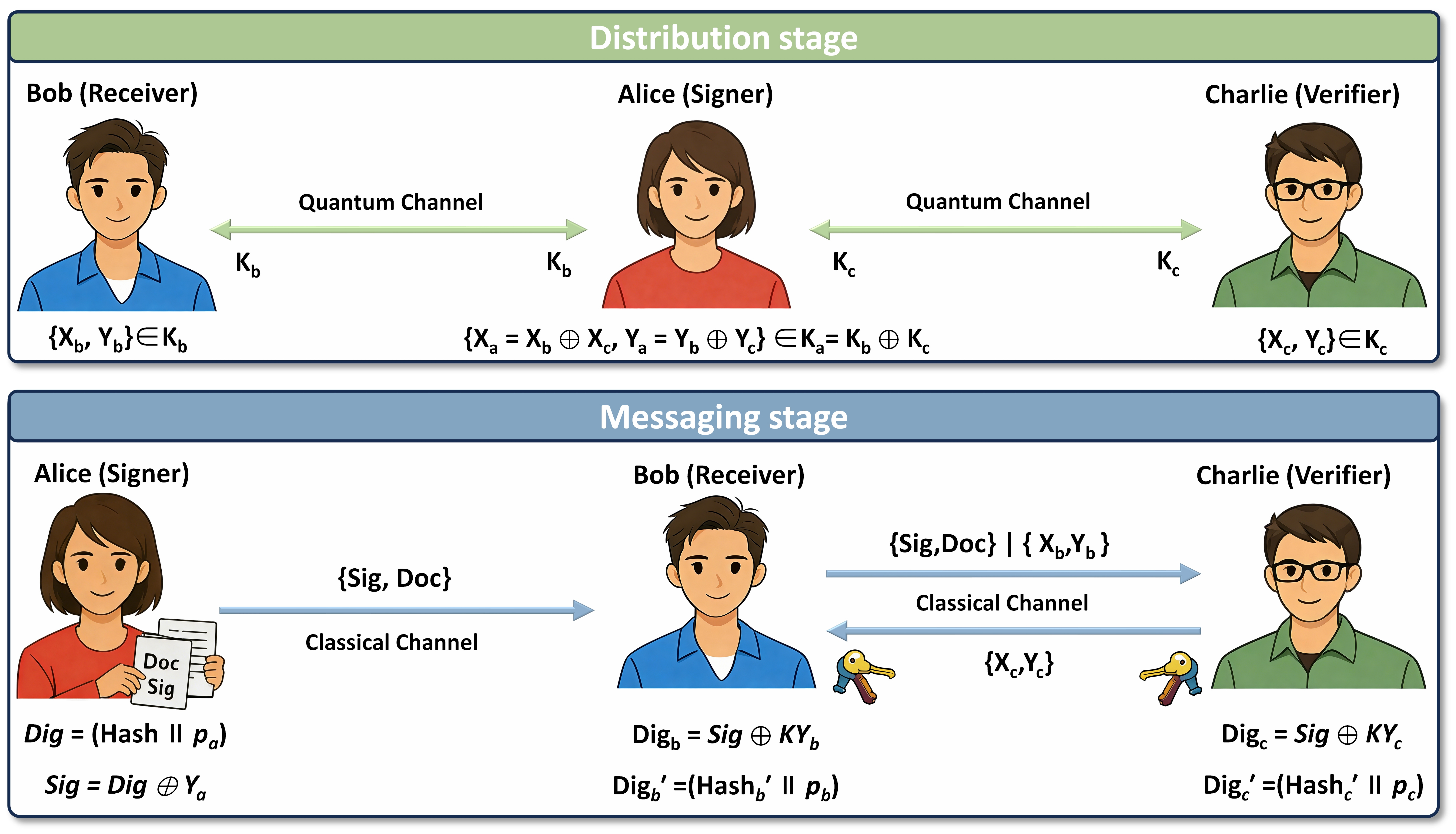}
		\end{tabular}
	\end{center}
	\caption {Schematic of the OTUH-QDS protocol. The protocol consists of two stages: the distribution stage and the messaging stage, involving three parties: Alice (Signer), Bob (Receiver), and Charlie (Verifier). In the distribution stage, Bob and Charlie independently perform the one-decoy-state BB84 QKD protocol with Alice to generate correlated key strings $K_b$ and $K_c$. Alice then computes her secret key $K_a = K_b \oplus K_c$, establishing the three-party key correlations $X_a = X_b \oplus X_c$ and $Y_a = Y_b \oplus Y_c$. In the messaging stage, Alice signs an $m$-bit document $\mathrm{Doc}$: she first computes the hash value $\mathit{Hash} = H_{n \times m} \cdot \mathit{Doc}$, concatenates it with a random bit string $p_a$ to form the digest $\mathit{Dig} = (\mathit{Hash} \,\|\, p_a)$, and encrypts it with her key $Y_a$ to produce the signature $\mathit{Sig} = (\mathit{Dig} \oplus Y_a)$. The signature and document $\{\mathit{Sig}, \mathit{Doc}\}$ are then sent to Bob. Bob forwards the message to Charlie and exchanges keys with him. Bob (Charlie) uses the reconstructed keys $KY_b$ and $KX_b$ ($KY_c$ and $KX_c$) to decrypt the signature, obtaining the expected digest $\mathit{Dig}_b$ ($\mathit{Dig}_c$), and independently computes the actual digest $\mathit{Dig}_b'$ ($\mathit{Dig}_c'$) via hashing. The signature is accepted as valid if and only if both Bob and Charlie verify the equality $\mathit{Dig}_b = \mathit{Dig}_b'$ and $\mathit{Dig}_c = \mathit{Dig}_c'$.}
	\label{Protocol}
\end{figure*}

To enhance the performance of the QDS system, we employ the OTUH-QDS protocol~\cite{yin2023experimental} as illustrated in Fig.~\ref{Protocol}. The protocol comprises two stages-distribution and messaging stage-and involves three participants: Alice (Signer), Bob (Receiver), and Charlie (Verifier).

\textbf{Distribution stage.} Bob (Charlie) independently perform a one-decoy-state QKD protocol with Alice to generate correlated key strings. The procedure is as follows: Bob (Charlie) randomly selects the $Z$ or $X$ basis to prepare one of the four BB84 polarization states, randomly modulates the pulse into a signal state (mean photon number $\mu$) or a decoy state (mean photon number $\nu$), and transmits it to Alice through an insecure quantum channel.
Alice randomly selects a measurement basis ($Z$ or $X$ basis) to detect the incoming pulses and records the detection events. Subsequently, Alice and Bob (or Charlie) announce their basis choices and intensity settings over an authenticated classical channel to perform basis sifting. Following post-processing steps, including parameter estimation, error correction, and privacy amplification, secret keys $K_b$ (between Alice and Bob) and $K_c$ (between Alice and Charlie) are generated.

Subsequently, Alice performs a bitwise XOR operation between the key strings $K_b$ and $K_c$ to generate her key $K_a = K_b \oplus K_c$, thus establishing the secret-sharing correlations among the three parties. When signing an $m$-bit document $\mathit{Doc}$, Alice randomly selects $3n$ bits from $K_a$ and splits them into two key strings of lengths $n$ and $2n$, denoted as $\{X_a, Y_a\}$. Alice then announces the positions of the selected key bits to Bob and Charlie over an authenticated classical channel. Based on this positional information, Bob and Charlie extract the corresponding key strings $\{X_b, Y_b\}$ and $\{X_c, Y_c\}$ from their own keys, satisfying the relations $X_a = X_b \oplus X_c$ and $Y_a = Y_b \oplus Y_c$.

\textbf{Messaging stage.} To sign an $m$-bit document $\mathit{Doc}$, the procedure proceeds as follows: First, Alice generates an $n$-bit random string $p_a$ using a local quantum random number generator. The random string $p_a$ is then used to construct an irreducible polynomial $p(x)$ of degree $n$.  Subsequently, Alice constructs an $n \times m$ Toeplitz hashing matrix $H_{n \times m}$. 
The matrix is generated using a linear feedback shift register (LFSR), with the secret key $X_a$ used as the initial vector and an irreducible polynomial $p(x)$ defining the feedback rule. The $m$-bit document $\mathit{Doc}$ is hashed using this matrix to obtain an $n$-bit hash value $\mathit{Hash} = H_{n \times m} \cdot \mathit{Doc}$. This hash value is concatenated with the random string $p_a$ to form a $2n$-bit digest $\mathit{Dig} = (\mathit{Hash} \parallel p_a)$. Alice encrypts the digest $\mathit{Dig}$ using a one-time pad (OTP) scheme with the key $Y_a$, generating a $2n$-bit signature $\mathit{Sig} = \mathit{Dig} \oplus Y_a$. The pair $\{\mathit{Sig}, \mathit{Doc}\}$ is transmitted to the receiver Bob via an authenticated classical channel.

Upon receiving $\{\mathit{Sig}, \mathit{Doc}\}$ from Alice, Bob forwards it along with his own key strings $\{X_b, Y_b\}$ to Charlie via an authenticated classical channel. Charlie subsequently transmits his key strings $\{X_c, Y_c\}$ to Bob through the same authenticated channel, completing the key exchange process. Bob then performs XOR operations to compute the combined keys $K{X_b} = X_b \oplus X_c$ and $K{Y_b} = Y_b \oplus Y_c$. Using $K{Y_b}$, Bob decrypts the signature to obtain the expected digest $\mathit{Dig}_b = \mathit{Sig} \oplus K{Y_b}$, from which he extracts the expected hash value $\mathit{Hash}_b$ and the random string $p_b$. Utilizing the initial vector ($K{X_b}$) and the string $p_b$, Bob reconstructs the Toeplitz hashing matrix $H'$ and applies it to the document $\mathit{Doc}$ to generate the actual hash value $\mathit{Hash}_b' = H' \cdot \mathit{Doc}$. The actual digest is formed as $\mathit{Dig}_b' = (\mathit{Hash}_b' \parallel p_b)$. The signature is accepted by Bob if $\mathit{Dig}_b = \mathit{Dig}_b'$; otherwise, it is rejected. If Bob accepts the signature, Charlie performs the same verification procedure to obtain the expected digest $\mathit{Dig}_c$ and the actual digest $\mathit{Dig}_c'$, and compares them to determine whether to accept the signature. 

Since each signature consumes $3n$ bits of secure key, the signature rate is directly determined by the secure key generation rate of the QKD system. We estimate the QKD secure key rate using a finite-key analysis method~\cite{rusca2018finite}. Thus, the signature rate is given by:  \begin{equation}
	\begin{split}
		R_{\mathrm{sig}} &= \Big[ s_{Z,0}^{L} + s_{Z,1}^{L} \big( 1 - h(e_{Z,1}^{\mathrm{ph}}) \big) 
		- \lambda_{\mathrm{EC}} \\
		&\quad - 6 \log_2 \Big( 19 / \epsilon_{\mathrm{sec}} \Big)
		- \log_2 \Big( 2 / \epsilon_{\mathrm{cor}} \Big) \Big] / (3n \cdot t),
	\end{split}
\end{equation}
where $t$ is the total data accumulation time; $s_{Z,0}^{L}$ and $s_{Z,1}^{L}$ are the lower bounds on the detection counts of vacuum and single-photon events in the $Z$ basis, respectively; $e_{Z,1}^{\mathrm{ph},U}$ is the upper bound on the phase error rate of single photon events; $\lambda_{\mathrm{EC}} = n_Z f_e h(e_Z)$ represents the information leakage during error correction. $f_e$ is the efficiency of the error correction code; and $\epsilon_{\mathrm{sec}}$ and $\epsilon_{\mathrm{cor}}$ denote the security and correctness parameters, respectively.

The security of the OTUH-QDS protocol is primarily characterized by forgery resistance, non-repudiation, and robustness. Specifically, the success probability of forgery attack is bounded by 
\(\epsilon_{\rm forge} \le m/2^{\,n-1}\), 
where \(m\) is the document length in bits and \(n\) is the hash output length. The success probability of repudiation attack is \(\epsilon_{\rm rep} = 0\). For robustness, the protocol's execution failure probability is determined by the error correction failure rate, with \(\epsilon_{\rm rob} = 2 \epsilon_{\rm cor}\). Consequently, the overall security parameter of the protocol is given by 
\(\epsilon_{\rm QDS} = \max(\epsilon_{\rm forge}, \epsilon_{\rm rep}, \epsilon_{\rm rob})\).
Further detail of the security analysis is provided in Supplementary Information Section 1.

\begin{figure*}[!t]
	\begin{center}
		\begin{tabular}{c}
			\includegraphics[height=7cm]{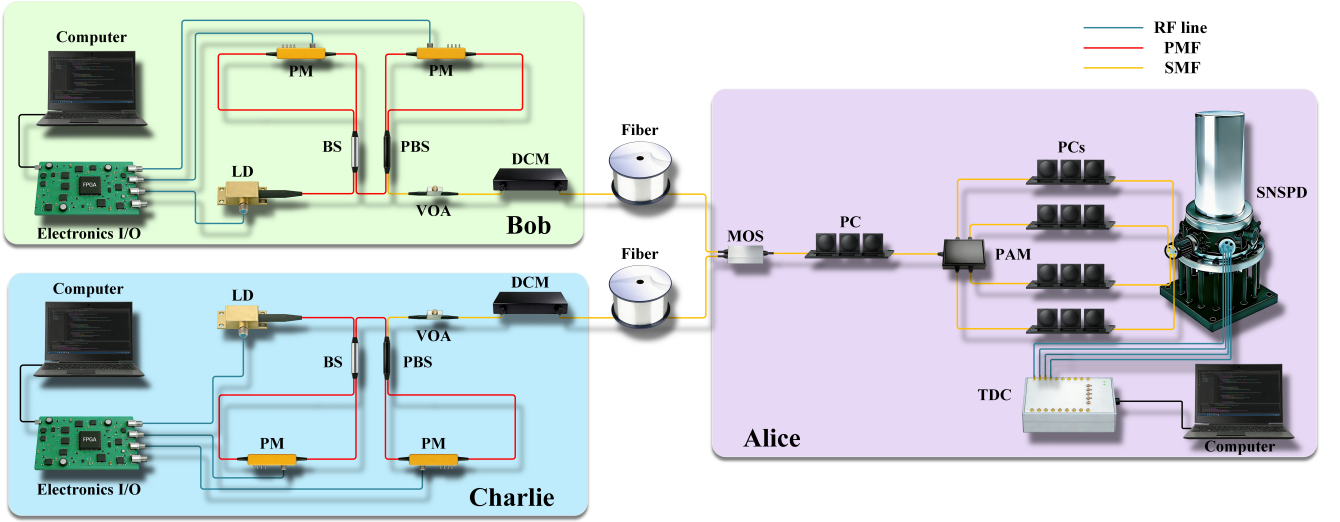}
		\end{tabular}
	\end{center}
	\caption {Schematic of the experimental setup of the QDS system. 
		The system consists of two independent transmitters (Bob and Charlie) and one receiver (Alice), with all nodes interconnected by commercial standard single-mode fibers. Each transmitter (Bob/Charlie) comprises control computer, FPGA board (Electronics I/O), laser diode (LD), phase modulator (PM), beam splitter (BS), polarization beam splitter (PBS), variable optical attenuator (VOA), and dispersion compensation module (DCM), which together enable quantum-state preparation, encoding, and transmission. The receiver (Alice) includes polarization controller (PC), polarization analysis module (PAM), superconducting-nanowire single-photon detector (SNSPD), time-to-digital converter (TDC), and computer for quantum-state measurement and data acquisition. Optical modules are interconnected via single-mode fiber (SMF) and polarization-maintaining fiber (PMF), while electrical signals are delivered through RF lines. The mechanical optical switch (MOS) is shown only for clarity of the system architecture and is not present in the actual experiment. } 
	\label{Setup}
\end{figure*}

\section{SETUP}

Figure.~\ref{Setup} shows the schematic of our experimental setup. We construct a QDS system comprising three nodes: Alice, Bob, and Charlie.

At the transmitting Bob (Charlie) side, phase-randomized optical pulses with a wavelength of 1549.48 nm and a repetition rate of 1.25 GHz are generated by a distributed feedback laser (signal laser, modelled LD-PD INC LP-ML1001A-55-FA), which is driven by a self-developed field-programmable gate array (FPGA) board. The generated pulses are then transmitted via a polarization-maintaining fiber (PMF) to a Sagnac interferometer-based intensity modulator (Sagnac IM) for intensity modulation. Within the Sagnac IM, each optical pulse is split into clockwise and counterclockwise propagating beams at a 50:50 polarization-maintaining beam splitter (PMBS). A lithium niobate phase modulator (PM) modulates the phase of one beam while leaving the other unaltered. The two beams subsequently recombine and interfere at the BS. By applying two distinct voltage amplitudes precisely to the PM via the self-developed FPGA board, optical pulses at two distinct intensity levels are generated for the one-decoy-state BB84 protocol~\cite{rusca2018finite}. Sagnac IM exhibits inherent stability due to the identical optical paths traversed by the two counter-propagating beams, effectively suppressing pattern-effect and thereby enhancing the system's security~\cite{yoshino2018quantum}.

Subsequently, the optical pulses are transmitted via polarization-maintaining fibers to a Sagnac interferometer--based polarization modulator (Sagnac POL) for preparing the four polarization states required by the BB84 protocol~\cite{1984-BB84}. 
Within the Sagnac POL, the optical pulses are split into two orthogonally polarized beams at a polarization beam splitter (PBS) and then propagate along the loop structure in the clockwise and counterclockwise directions, respectively. 
Within the loop, a phase shift is applied by the PM to one of the beams (e.g., the clockwise one), thereby creating a specific relative phase difference between the two counter-propagating beams. 
The two beams subsequently interfere upon recombination at the PBS. 
As a result, the polarization state of the optical pulses at the output of the Sagnac POL can be expressed as 
\begin{equation}
	\lvert \psi \rangle = \frac{1}{\sqrt{2}} \left( \lvert H \rangle + e^{i\theta} \lvert V \rangle \right).
\end{equation}

Following the polarization encoding, the optical pulses are attenuated to the single-photon level by variable optical attenuator (VOA). 
To compensate for chromatic dispersion introduced during fiber transmission, a dispersion compensation module (DCM) is inserted before the quantum channel, effectively suppressing the temporal broadening of high-speed optical pulses over long-distance transmission. The insertion loss of the DCM is included in the overall attenuation at the transmitting Bob (or Charlie) side. 
The optical pulses are then transmitted to Alice through the quantum channel, which consists of standard commercial single-mode fiber (G.652.D) with an attenuation coefficient of approximately $0.2~\mathrm{dB/km}$.

To prepare the four BB84 polarization states, the self-developed FPGA board randomly selects and applies four distinct voltage levels to the PM, corresponding to phase shifts $\theta \in \{0, \pi/2, \pi, 3\pi/2\}$. 
Here, $\theta \in \{0, \pi\}$ corresponds to the $Z$ basis, while $\theta \in \{\pi/2, 3\pi/2\}$ corresponds to the $X$ basis.

\begin{figure*}[!t]
	\centering  
	\includegraphics[height=12.5cm]{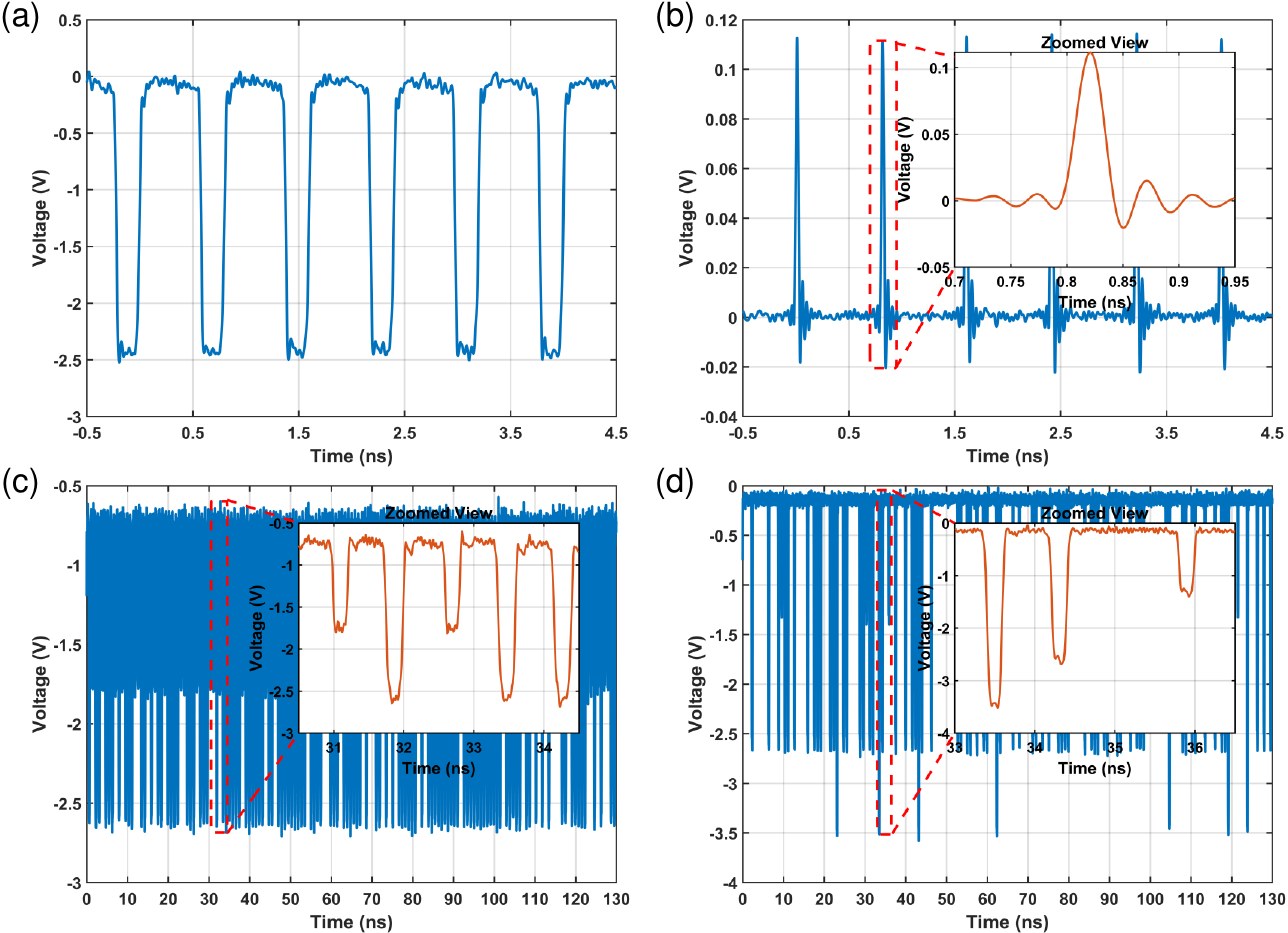}
	\caption{Electrical and optical signals for high-speed quantum-state preparation at a \SI{1.25}{GHz} repetition rate. All signals are recorded using a \SI{25}{GHz}-bandwidth oscilloscope. (a) Laser drive electrical signal. This signal is generated by a self-developed FPGA board, with a repetition rate of \SI{1.25}{GHz} and a pulse width of approximately \SI{200}{ps}, and is used for gain-switching operation of a DFB semiconductor laser; (b) Output optical pulse from the laser. The pulse is produced by gain-switching the DFB laser driven by the electrical signal in (a), exhibiting an FWHM of approximately \SI{30}{ps}. It is measured by a high-speed photodiode with \SI{20}{GHz} bandwidth; (c) Intensity modulator drive electrical signal. This signal is used for preparing signal and decoy states, with a pulse width of about \SI{200}{ps}. The waveform shown here is after \SI{6}{dB} broadband attenuation; (d) Polarization modulator drive electrical signal. This signal is employed for encoding the four polarization states ($|H\rangle$, $|V\rangle$, $|D\rangle$, $|A\rangle$) of the BB84 protocol by switching among four distinct voltage levels. Its pulse width is \SI{200}{ps}. The displayed signal is after \SI{6}{dB} broadband attenuation.}
	\label{waveform1}
\end{figure*}

At the receiving side, Alice detects the quantum signal from Bob (Charlie). First, a polarization controller (PC) is used to calibrate the polarization basis, such that the incident polarization states are properly aligned with the $X$ and $Z$ bases of the polarization analysis module (PAM). After polarization calibration, the optical pulses enter the PAM for polarization analysis, and the output photons are directed to four commercial SNSPDs for detection. The arrival times of the detected photons are recorded by a time-to-digital converter (TDC), providing the timing information required for subsequent data processing.

\section{PERFORMANCE CHARACTERIZATION}

For high-rate QDS experiments, significant technical challenges arise in the high-speed preparation and detection of quantum states. 
First, stable generation of ultrashort optical pulses is required to avoid temporal overlap between adjacent pulses during encoding and inter-pulse crosstalk during detection. 
To this end, we use a self-developed FPGA board to generate 200-ps electrical pulses at a repetition rate of 1.25~GHz to gain-switch a distributed-feedback (DFB) semiconductor laser, exploiting the laser's relaxation-oscillation dynamics to produce ultrashort optical pulses with a full width at half maximum (FWHM) of 30~ps. The driving signal of the laser and the resulting output optical pulses are shown in Figs.~\ref{waveform1}(a) and~\ref{waveform1}(b)

Second, to achieve stable quantum-state modulation, the temporal separation between the clockwise and counterclockwise propagating pulses in the Sagnac interferometer is precisely set to 400~ps, while the PM is driven by 200-ps-wide electrical pulses from the self-developed FPGA board. 
This configuration effectively suppresses modulation crosstalk that would arise if a single electrical pulse acted simultaneously on both propagating beams within the PM. 
The electrical driving signals for intensity and polarization modulation are shown in Figs~\ref{waveform1}(c) and~\ref{waveform1}(d). Performance characterization results shows that the maximum extinction ratios of the intensity modulators at the Bob and Charlie transmitters reach 35.98~dB and 34.63~dB, respectively. For the four polarization states ($\lvert H\rangle$, $\lvert V\rangle$, $\lvert D\rangle$, $\lvert A\rangle$) required for polarization encoding, the measured extinction ratios are 27.70~dB, 26.34~dB, 26.07~dB, and 20.72~dB at Bob, and 27.30~dB, 26.39~dB, 22.25~dB, and 21.42~dB at Charlie, respectively. These results demonstrate that the system maintains high-fidelity quantum state encoding even at a repetition rate of 1.25~GHz.

For quantum state measurements, we employ SNSPD based on the structure of a dielectric mirror (SiO$_2$/Ta$_2$O$_5$ film stack) and sandwiched twin-layer NbN nanowires with an intermediate SiO$_2$ insulating layer, as proposed by Hu et al.~\cite{hu2020detecting} The detectors achieve a system detection efficiency exceeding 70\% at 1550~nm wavelength, with a timing jitter below 40~ps, a maximum count rate of 10~MHz, and a dark count rate maintained at 25~Hz. 
Their high efficiency, high count rate, and low jitter characteristics provide critical support for quantum state measurement. 

In this experiment, to balance count-rate linearity and long-term system stability under high-repetition-rate operation, the system detection efficiency is set to approximately $60\%$. Benefiting from the optimized optical modulation and detection performance described above, the intrinsic quantum bit error rate (QBER) of the system is measured to be $2.14\%$ on the Bob--Alice link and $1.91\%$ on the Charlie--Alice link. These results confirm the high reliability of the system under high-speed operation.

\section{RESULTS}

Using the described set-up, we systematically evaluated the practical performance of the QDS system under different transmission distances in a laboratory environment. In the OTUH-QDS protocol~\cite{yin2023experimental}, the protocol robustness satisfies $\epsilon_{\rm rob}=2\epsilon_{\rm cor}$. In our experiment, $\epsilon_{\rm cor}$ is fixed at $10^{-15}$, such that the overall system security is primarily constrained by the unforgeability. For a document size of $m = 1$~Mbit, we set the hash value 
$n = 51$ bits, yielding a forgery probability of $\epsilon_{\rm forge}\approx 8.88\times 10^{-10}$, which is well below the QKD security parameter $\epsilon_{\mathrm{sec}} = 10^{-9}$. Therefore, the overall security of the system is given by $\epsilon = 10^{-9}$.

To achieve optimal performance while satisfying the aforementioned security conditions, we conduct comprehensive optimization of the implementation parameters at various transmission distances\textemdash including signal/decoy-state mean photon numbers, sending probabilities, and basis selection probabilities. For instance, in the 100~km demonstration, the signal-state and decoy-state mean photon numbers for Bob (Charlie) are selected as $\mu = 0.393$ (0.418) and $\nu = 0.108$ (0.115), respectively. The probabilities for sending the signal state $\mu$ and selecting the Z-basis are set as $P_\mu = 0.737$ (0.745) and $P_z = 0.926$ (0.930). Based on the above parameter optimization and long-term stable experimental data acquisition, we obtain the effective signature rates at different transmission distances: over commercial single-mode fibers of 75~km, 100~km, 150~km, 200~km, and 250~km (losses of 14.24~dB, 19.59~dB, 29.00~dB, 38.98~dB, and 49.05~dB, respectively), the system achieves signature rates of 5186.80~tps, 1500.20~tps, 160.83~tps, 14.92~tps, and 1.25~tps. Detailed experimental data are provided in Supplementary Information Section 2.

\begin{figure}[btp]
	\centering
	\includegraphics[width=0.95\columnwidth]{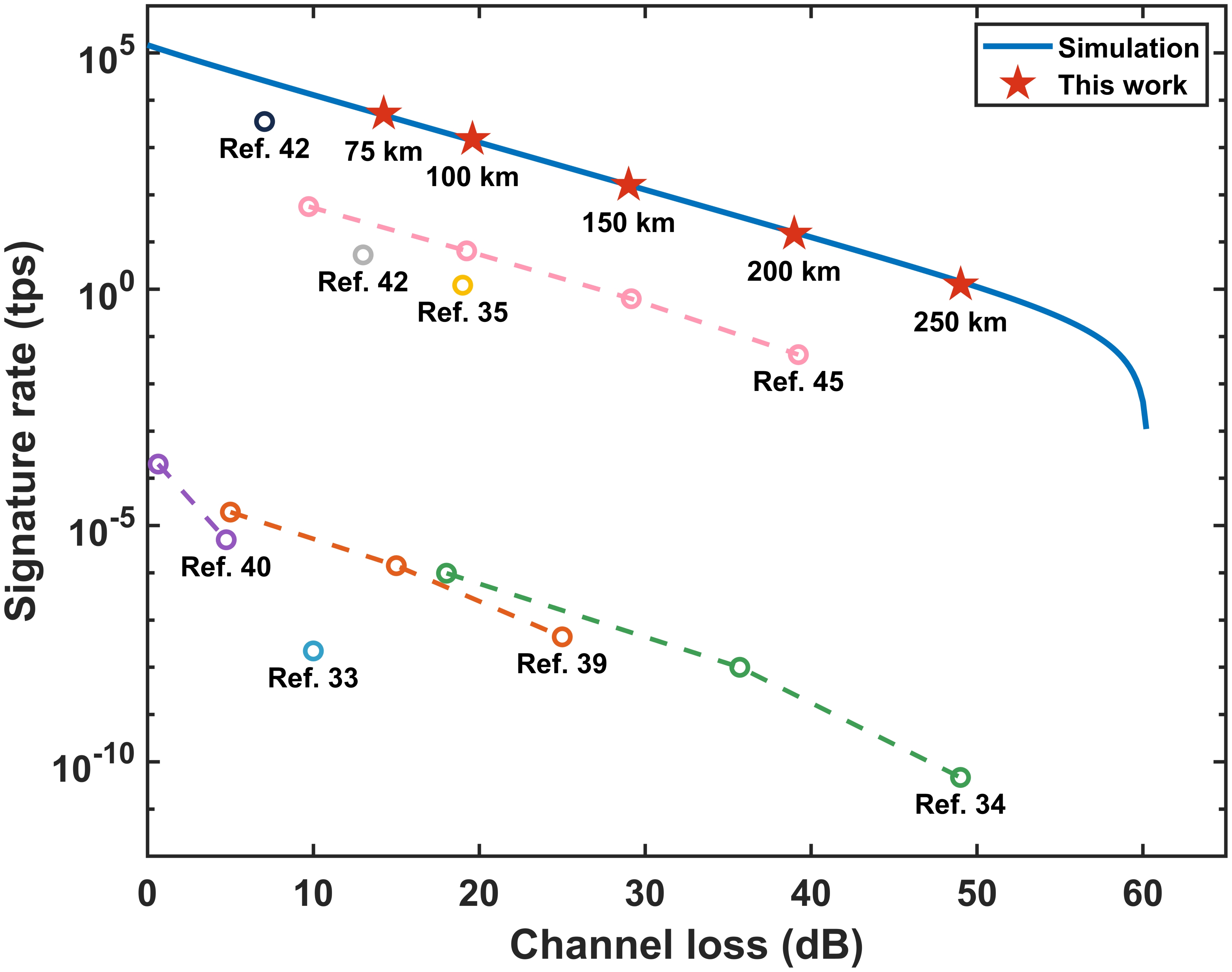}
	\caption{Signature rates with different transmission loss. The blue solid line shows the theoretical simulation results obtained from the experimental system parameters, while the red pentagrams denote the signature rates achieved in this work over commercial standard single-mode fiber at 75 km, 100 km, 150 km, 200 km and 250 km. The circles correspond to the signature rate results from state-of-the-art QDS experiments (blue circle~\cite{roberts2017experimental}, orange circle~\cite{an2018practical}, green circle~\cite{ding2020280}, purple circle~\cite{richter2021agile}, yellow circle~\cite{yin2023experimental}, pink circle~\cite{du2025chip}, black circle~\cite{lu2025fully}, and gray circle~\cite{lu2025fully}), where data points from the same reference under different channel losses are connected by dashed lines.}
	\label{Result}
\end{figure}

To highlight the progress enabled by our results, we compare the achieved signature rates with those reported in state-of-the-art QDS experiments, as summarized in Fig.~\ref{Result} and Table~\ref{table_experi_compare}.

\begin{table*}[!t]
	\caption{Comparison of state-of-the-art QDS experiments. $\epsilon$ is the security parameter, and $R_{\mathrm{sig}}$ represents
		the signature rate. }
	
	\setlength\tabcolsep{4pt}
	\renewcommand\arraystretch{1.3}
	
	\noindent
	\begin{tabular*}{\linewidth}{@{\extracolsep{\fill}}lcccccccc}
		\hline\hline  
		References & Protocol & Clock rate (MHz)& Distance (km) & Loss (dB) & Document size (bit) & $\epsilon$ & $R_{\mathrm{sig}}$ (tps) \\ 
		\hline  
		
		Roberts et al.~\cite{roberts2017experimental} & MDI  & 1000  & 50 & 10.00  \textsuperscript{a}  & 1  & $1\times10^{-10}$ & $2.22\times10^{-2}$  \\
		\hline
		
		An et al.~\cite{an2018practical}           & BB84 & 1000  & 125 & 25.00 \textsuperscript{a} & 1  & $1\times10^{-10}$ & $4.41\times10^{-2}$ \\
		\hline
		
		\multirow{3}{*}{Ding et al.~\cite{ding2020280}}      
		& \multirow{3}{*}{BB84}
		& \multirow{3}{*}{50} 
		& 103  & 18.03 \textsuperscript{a}
		& \multirow{3}{*}{1}  
		& \multirow{3}{*}{$1\times10^{-5}$}  
		& 0.98 \\
		& & & 204 & 35.70 \textsuperscript{a} & & & 0.01 \\
		& & & 280 & 49.00 \textsuperscript{a} & & & $4.67\times10^{-5}$ \\
		\hline
		
		Richter et al.~\cite{richter2021agile} & CV   & 1000  & 20 & 4.75  & 1  & $1\times10^{-4}$  & 5 \\
		\hline
		
		Yin et al.~\cite{yin2023experimental}    & BB84 & 200 & 101 & 19.00 & $10^6$ & $1\times10^{-32}$ & 1.22 \\[2pt]
		\hline
		
		\multirow{2}{*}{Du et al.~\cite{du2025chip}}
		& \multirow{2}{*}{BB84}
		& \multirow{2}{*}{50 }
		& 100 & 19.24 
		& \multirow{2}{*}{$10^6$}
		& \multirow{2}{*}{$4.72\times10^{-8}$}
		& 6.50 \\[2pt]
		& & & 200 & 39.23 & & & $4.14\times10^{-2}$ \\
		\hline
		
		Lu et al.~\cite{lu2025fully}
		& BB84 & 1000 & 35.3 & 7.06 \textsuperscript{a} & $3.44\times10^{5}$ & - & 3545 \\
		\hline
		Lu et al.~\cite{lu2025fully}
		& MDI & 1000 & 65 & 13.00 \textsuperscript{a} & $3.44\times10^{5}$  & - & 5.34 \\
		\hline
		
		\multirow{4}{*}{Our work}
		& \multirow{4}{*}{BB84}
		& \multirow{4}{*}{1250}
		& 35.3 & 7.06
		& \multirow{4}{*}{$10^6$}
		& \multirow{4}{*}{$1\times10^{-9}$}
		& 25815.85 \textsuperscript{b} \\
		& & & 100 & 19.59 & & & 1500.20 \\
		& & & 200 & 38.98 & & & 14.92 \\
		& & & 250 & 49.05 & & & 1.25 
		\\ \hline\hline
		
	\end{tabular*}
	\vspace{6pt}
	\small
	\noindent
	\parbox[t]{\linewidth}{
		\raggedright
		\textit{Notes:}
		$^{a}$ Calculated from the loss coefficient and transmission distance reported in the original reference;
		$^{b}$ Simulated experiment;
		- Parameter not reported in the original reference.}
	\label{table_experi_compare}
\end{table*}

In terms of signature rate, our work achieves a marked breakthrough: under comparable channel-loss conditions, the signature rate exceed those of existing studies by at least two orders of magnitude. In particular, at a transmission distance of 75~km our system attains a signature rate of 5,186.80~tps, setting a new record for QDS experiments. This performance notably surpasses the rate of 3545~tps achieved by Lu et al.~\cite{lu2025fully} at a shorter distance of 35.3~km; despite a channel loss approximately 2.8 times higher in our system, the signature rate is increased by about 1.46 times. (We also conducted a numerical simulation based on the actual parameters of our experimental platform for a 35.3~km link, yielding a signature rate of 25,815.85~tps, which further confirms the performance potential of our system.)

In terms of tolerance to high channel loss, under approximately 49~dB channel loss, Ding et al.~\cite{li2019high} reported a signature rate of  $4.67\times10^{-5}$~tps for a 1-bit document over 280~km of ultra-low-loss fiber (loss of approximately 49 dB). In contrast, our system achieves a signature rate of 1.25~tps for a 1~Mbit document over 250~km of standard commercial fiber (measured total loss: 49.05~dB). While the document size is increased by a factor of one million, the signature rate is improved by over four orders of magnitude, unequivocally demonstrating the superior high-loss tolerance of our system.

\section{Conclusion}\label{Conclusion}

To overcome the critical limitations of low signature rates and insufficient channel-loss tolerance in practical QDS implementations, we propose and experimentally demonstrate a high-performance QDS system. The system combines intrinsically phase-stable polarization modulation based on a Sagnac interferometer, gigahertz-rate quantum state encoding techniques and low-timing-jitter SNSPDs. Together with the OTUH-QDS protocol, stable operation is achieved at a clock frequency of 1.25\,GHz, enabling the realization of a QDS experimental platform that simultaneously supports high signature rates and high tolerance to channel loss.

Compared to existing QDS demonstrations, our work improves the signature rate by at least two orders of magnitude under comparable channel loss conditions. In particular, a signature rate of 5186.80\,tps is achieved over 75\,km of standard single-mode fiber, setting a new performance record among reported QDS experiments. Even under a total channel loss as high as 49.05\,dB (corresponding to 250\,km of fiber), the system maintains an effective signature rate of about 1.25\,tps, fully demonstrating its excellent performance. Looking ahead, the combination of high signature rates and high channel loss tolerance realized in this work provides a viable technological foundation for the practical deployment of QDS in future quantum communication networks, and lays important groundwork for its applications in scenarios such as quantum timestamping~\cite{li2025information}, quantum blockchain~\cite{weng2023beating}, and quantum e-commerce~\cite{cao2024experimental}.

\section*{ACKNOWLEDGMENTS}
This study was supported by the National Natural Science Foundation of China (Nos. 62171144, 62031024, and  11865004), Guangxi Science Foundation (2025GXNSFAA069137),  Guangdong Basic and Applied Basic Research Foundation (2024B1515120030), and Bagui Scholars Programme (W.X.-G., GXR-6BG2424001).

\textit{Note added.—}
Recently, we became aware of a similar work by Zhang \textit{et al.}~\cite{zhang2026experimentaldemonstrationtwinfieldquantum}, which reports a 1.25~GHz QDS experiment based on the twin-field protocol. In this work, we experimentally demonstrate a 1.25~GHz QDS system based on the BB84 protocol. 

\begin{table*}[htp]
	\centering
	\setlength{\tabcolsep}{6.5pt}
	\renewcommand{\arraystretch}{1.7}
	\caption{Experimental parameters and performance for the A--B (Alice--Bob) and A--C (Alice--Charlie) links at different transmission distances. $T$ denotes the link transmission distance; Loss represents the optical loss of the fiber link; $\mu$ and $\nu$ are the intensities of the signal and decoy states, respectively; $P_\mu$ and $P_\nu$ are the probabilities of sending the signal and decoy states; $P_Z$ and $P_X$ denote the basis selection probabilities for the $Z$ and $X$ bases, respectively. $n_{Z,\mu}$ and $n_{Z,\nu}$ are the total detection events of the signal and decoy states in the $Z$ basis, and $m_{Z,\mu}$ and $m_{Z,\nu}$ are the corresponding error counts; $n_{X,\mu}$ and $n_{X,\nu}$ are the total detection events of the signal and decoy states in the $X$ basis, and $m_{X,\mu}$ and $m_{X,\nu}$ are the corresponding error counts; $n_Z$ is the total number of accumulated detection events in the $Z$ basis after basis reconciliation; $t$ is the time required to accumulate $n_Z$; $s_{Z,1}^L$ denotes the lower bound of single-photon detection events in the $Z$ basis; $E_Z$ is the quantum bit error rate in the $Z$ basis; $\phi_Z^U$ represents the upper bound of the phase error rate in the $Z$ basis; $\lambda_{\mathit{EC}}$ is the number of bits leaked during error correction; SKR is the secure key rate; $R_{\mathit{sig}}$ is the signature rate.}
	
	\begin{tabular}{lcccccccccc}
		\toprule
		$T$ (km)
		& \multicolumn{2}{c}{75}
		& \multicolumn{2}{c}{100}
		& \multicolumn{2}{c}{150}
		& \multicolumn{2}{c}{200}
		& \multicolumn{2}{c}{250} \\
		\cmidrule(lr){2-3}\cmidrule(lr){4-5}\cmidrule(lr){6-7}
		\cmidrule(lr){8-9}\cmidrule(lr){10-11}
		Link
		& A--B & A--C
		& A--B & A--C
		& A--B & A--C
		& A--B & A--C
		& A--B & A--C \\
		\midrule
		
		Loss (dB)
		& 14.24 & 14.24
		& 19.59 & 19.59
		& 29.00 & 29.00
		& 38.98 & 38.98
		& 49.05 & 49.05 \\
		
		$\mu$
		& 0.396 & 0.420
		& 0.393 & 0.418
		& 0.393 & 0.417
		& 0.393 & 0.418
		& 0.392 & 0.417 \\
		
		$\nu$
		& 0.109 & 0.117
		& 0.108 & 0.116
		& 0.108 & 0.115
		& 0.108 & 0.115
		& 0.107 & 0.113 \\
		
		$P_{\mu}$
		& 0.736 & 0.745
		& 0.737 & 0.745
		& 0.738 & 0.746
		& 0.737 & 0.745
		& 0.736 & 0.745 \\
		
		$P_{\nu}$
		& 0.264 & 0.255
		& 0.263 & 0.255
		& 0.262 & 0.254
		& 0.263 & 0.255
		& 0.264 & 0.255 \\
		
		$P_Z$
		& 0.927 & 0.930
		& 0.926 & 0.930
		& 0.926 & 0.929
		& 0.925 & 0.929
		& 0.920 & 0.923 \\
		
		$P_X$
		& 0.073 & 0.070
		& 0.074 & 0.070
		& 0.074 & 0.071
		& 0.075 & 0.071
		& 0.080 & 0.077 \\
		
		$n_{Z,\mu}$
		& 9022934 & 9135866
		& 9108386 & 9135586
		& 9097920 & 9148512
		& 9088230 & 9152591
		& 9056811 & 9132025 \\
		
		$m_{Z,\mu}$
		& 188837 & 145890
		& 158161 & 150499
		& 149331 & 171872
		& 192606 & 167624
		& 223941 & 173504 \\
		
		$n_{X,\mu}$
		& 732612 & 682682
		& 707459 & 656318
		& 726566 & 694262
		& 746922 & 680266
		& 794800 & 745538 \\
		
		$m_{X,\mu}$
		& 24623 & 16155
		& 16989 & 19779
		& 21574 & 15167
		& 18152 & 14238
		& 25157 & 18129 \\
		
		$n_{Z,\nu}$
		& 977066 & 864134
		& 891614 & 864414
		& 902080 & 851488
		& 911770 & 847409
		& 943189 & 867975 \\
		
		$m_{Z,\nu}$
		& 32560 & 25158
		& 22969 & 22632
		& 18543 & 20240
		& 22670 & 19512
		& 36012 & 31488 \\
		
		$n_{X,\nu}$
		& 79535 & 74207
		& 75523 & 70010
		& 73413 & 69000
		& 69223 & 66327
		& 78671 & 68890 \\
		
		$m_{X,\nu}$
		& 4414 & 3713
		& 2619 & 3096
		& 2406 & 1722
		& 1806 & 1741
		& 3274 & 2530 \\
		
		$n_Z$
		& $10^7$ & $10^7$
		& $10^7$ & $10^7$
		& $10^7$ & $10^7$
		& $10^7$ & $10^7$
		& $10^7$ & $10^7$ \\
		
		$t$ (s)
		& 3.48 & 3.29
		& 12.81 & 11.21
		& 107.3 & 98.7
		& 1037.2 & 975.0
		& 10683.9 & 10804.3 \\
		
		$s^{l}_{Z,1}$
		& 6746999 & 5749700 
		& 6018444 & 5826217 
		& 6241109 & 5821895 
		& 6246379 & 5788676 
		& 6396592  & 5852624  \\
		
		$E_Z$ (\%)
		& 2.214 & 1.711
		& 1.811 & 1.731
		& 1.679 & 1.921
		& 2.153 & 1.871
		& 2.599 & 2.050 \\
		
		$\phi_Z^{u}$
		& 0.0598 & 0.0342
		& 0.0438 & 0.0546
		& 0.0653 & 0.0471
		& 0.0638 & 0.0456
		& 0.0696 & 0.0559 \\
		
		$\lambda_{\rm EC}$
		& 1778180 & 1448383
		& 1516239 & 1462483
		& 1426814 & 1589066
		& 1739231 & 1556204
		& 2017154 & 1673118 \\
		
		SKR
		& 793581 & 931235
		& 229531 & 230299
		& 24607 & 26723
		& 2283 & 2754
		& 192 & 218 \\

		$R_{\mathit{sig}}$ (tps)
		& \multicolumn{2}{c}{5186.80}
		& \multicolumn{2}{c}{1500.20}
		& \multicolumn{2}{c}{160.83}
		& \multicolumn{2}{c}{14.92}
		& \multicolumn{2}{c}{1.25} \\
		
		\bottomrule
	\end{tabular}
	\label{table_rawdate}
\end{table*}

 \appendix

 \section{Security Analysis of OUTH-QDS}\label{appendix A}
The security of the one-time universal hashing quantum digital signature (OTUH-QDS) protocol~\cite{yin2023experimental} primarily encompasses forgery resistance, non-repudiation, and robustness.

 \textbf{1. Forgery}

 Forgery attack refers to an adversary (Bob) attempting to construct a fraudulent message--signature pair $\{M', \mathit{Sig}'\}$ (with $M' \neq M$) that will be accepted by the verifier (Charlie), without knowledge of Alice's private keys $\{X_a, Y_a\}$ and the random string $p_a$. In the protocol, the signature $\mathit{Sig}$ is encrypted using the one-time key $Y_a = Y_b \oplus Y_c$, and the hash function used to generate the digest is uniquely determined by $\{X_a, p_a\}$. Thus, the uncertainty in $Y_a$ and $p_a$ entirely originates from the unknown keys of the other party $\{X_c, Y_c\}$. Within the security framework of the one-time pad and the OTUH, the adversary's optimal forgery strategy is equivalent to correctly guessing the specific hash function used for the current signature.

 For LFSR-based Toeplitz hashing, the collision probability has a known upper bound~\cite{krawczyk1994lfsr}, and thus the success probability of forgery can be bounded as:
 \begin{equation}
 	\epsilon_{\mathit{forge}} \leq \frac{m}{2^{n-1}},
 \end{equation}
 where $m$ is the size of the document and $n$ is the length of the hash value

 \textbf{2. Repudiation}

 Repudiation attack refers to a scenario where the signer Alice, after generating a signature, attempts to cause honest Bob and Charlie to reach divergent verification conclusions for the same message -signature pair. In the verification phase of the protocol, Bob and Charlie first exchange their respective key shares via an authenticated classical channel and compute the combined keys
 $KX = X_b \oplus X_c$ and $KY = Y_b \oplus Y_c .$
 The authenticated channel guarantees the integrity and authenticity of the exchanged information, ensuring that both parties obtain identical $KX$ and $KY$. Consequently, they decrypt the signature to derive the same expected digest and reconstruct a consistent hash function. Under the condition of honest protocol execution, Bob and Charlie are thus bound to achieve identical verification results for the same signature, rendering the repudiation attack impossible. The success probability of such an attack is rigorously bounded as
 $
 \epsilon_{\mathit{rep}} = 0 .
 $

 \textbf{3. Robustness}

 Robustness refers to the probability that a signature is correctly accepted when all participants execute the protocol honestly. The potential failure of the protocol primarily stems from error correction failures during the key distribution stage, with a failure probability denoted as $\epsilon_{\mathit{cor}}$, the robustness failure probability satisfies
 $
 \epsilon_{\mathit{rob}} = 2 \epsilon_{\mathit{cor}} .
 $
 
 In conclusion, the overall security parameter of the OTUH QDS protocol is given by 
 \begin{equation}
 	\epsilon_{\mathit{QDS}} = \max \left(\epsilon_{\mathit{forge}}, \epsilon_{\mathit{rep}}, 
 	\epsilon_{\mathit{rob}} \right)
 \end{equation}

 \section{Detailed Experimental Results}\label{appendix B}
 Table~\ref{table_rawdate}. shows the detailed experimental results.
 %


\bibliography{reference}

\bibliographystyle{naturemag}


%

\end{document}